\documentstyle [12pt]{article}
\begin{document}
\begin{titlepage}
\begin{flushright}
PUPT-1301 \\
December, 1991
\end{flushright}
\vspace{20mm}
\begin{center}
\huge{Symmetries and Special States}
\end{center}
\vspace{1 mm}
\begin{center}
\huge{in}
\end{center}
\vspace{1 mm}
\begin{center}
\huge{Two Dimensional String Theory}
\end{center}
\vspace{10mm}
\begin{center}
Ulf H. Danielsson\  \footnote{Work supported by a Harold W. Dodds
Fellowship} \\
{\em Joseph Henry Laboratories} \\
{\em Princeton University} \\
{\em Princeton, NJ 08544} \\
{\em USA}
\end{center}
\begin{center}
{\large Abstract}

\small
We use the $W_{\infty}$ symmetry of $c=1$ quantum gravity to compute
matrix model special state correlation functions. The results are
compared, and found to agree, with expectations from the Liouville
model.
\end{center}
\end{titlepage}

\section{Introduction}

The last few years have seen tremendous developements in the
understanding of two dimensional quantum gravity and therefore non
critical string theory. The first success came from field theory,
\cite{DDK}, where the case with matter of central charge $c \leq 1$
coupled to quantum gravity was solved. Later these theories were also
solved by the powerful matrix models, both for $c<1$ \cite{m1}, and for
$c=1$ \cite{m2}. They allow for exact, non
perturbative solutions where one sums over all genus. This is an
important improvement over field theory where higher genus contributions
are extremely difficult to calculate. Most results have therefore been
limited to the sphere or in some cases the torus.

These models are important for two reasons. First they give solutions of
two dimensional quantum gravity and can serve as toy models for higher
dimensional theories. Indeed, the matrix model provides directly the
exact Wheeler de Witt equation summed over all genus, \cite{mo}. Clearly an
important object for further study. Second, and perhaps even more
interesting, they describe non critical string theories. The most
physical example is clearly the case $c=1$. It is now commonplace to
identify the Liouville mode as an extra dimension, \cite{liou},
and thereby obtaining
a theory of strings moving in a two dimensional target space. Naively
one would expect this theory to be very simple, just a single massless
scalar particle, usually referred to as the tachyon. This can be argued
by choosing lightcone gauge. Fortunately, this is not the whole story.
Instead there are remnants of the massive excited string modes present
for certain discrete values of the momentum, \cite{mo,art2,new,poly}.
They are usually called
special or discrete states. Clearly it is very interesting to study
these extra states if one wants to learn about truly stringy phenomena.

The special states have been the object for several recent studies. Both
using  Liouville theory and matrix models. An important discovery has
been a huge set of symmetries.
These symmetries obey a $W_{\infty}$ algebra, which can be thought of as a
generalization of a Virasoro algebra. This has been shown in the
Liouville theory with two different methods. In \cite{klpol} the
$W_{\infty}$ was found by explicitly calculating the operator product
expansion of the special states. An important tool for doing so is the
usual $SU(2)$ symmetry known for a long time.
In \cite{wit} the symmetry was instead found from the construction of the
ground
ring. This gives a representation of the current algebra with the
currents acting on a set of
ghost number
zero, spin zero fields. These fields can be shown to be primary,
\cite{zuck}, and correspond to special states in addition to the
standard ones. The meaning of these and other special states of non
standard ghost numbers has as yet not been fully clarified.
{}From the matrix model point of view,
the emergence of the symmetry has been more gradual \cite{wit,mose,min,ant}.
In \cite{wit} it was however clearly realized that the $W_{\infty}$
simply is generated by the matrix eigenvalue and its conjugated momentum
through the Poisson bracketts.

In this work we will use the $W_{\infty}$ to
study the special operator correlation functions. Notations and
conventions will be much the same as in a previous paper, \cite{art2}.
We will be able to obtain many of the results in \cite{art2} using the
simplifications the $W_{\infty}$ symmetry provides.

In section 2 we
make some initial comments on Ward identities and symmetries relating
to the results obtained in \cite{art2}. We will also make some comments
on possible generalizations to non harmonic matrix model potentials based on a
generalized Wheeler de Witt equation. In section 3 we calculate some
matrix model special state correlation functions using the $W_{\infty}$
symmetry. We also make some comments on how to identify the counterparts
of the Liouville model special states. Section 4 gives some
illustrations of the structure one encounters at higher genus. Finally,
section 5 is devoted to a comparison with Liouville theory. Although the
success of the matrix model and its agreement with Liouville theory
hardly is in doubt, it is important to make the connection as explicit
as possible. In particular, it is so far not clear how to explicitly
extract the space time structure from the matrix model. One would like
to be able to study nontrivial space times like the recently discovered
black hole solution \cite{svart}. We will not be able to adress this
question here, but we will be able to make a comparison between our
matrix model results and some Liouville theory expectations. In the
cases which we will examine we will find perfect agreement.

\section{Ward Identities and Symmetries}

Let us now consider the matrix model and its special states and
operators. The matrix model represents the string theory Riemann
surfaces by Feynman diagrams of interacting
matrix variables which triangulate the surfaces. In the uncompactified case, or
at least for large
enough radius, we can simply integrate out most of the degrees of
freedom. The only remaining will be the matrix eigenvalues. They will
then behave as non interacting fermions in the matrix model potential.
Obvious candidates for special state correlation functions are then
correlation functions of powers of the matrix model eigenvalues. Such
objects were studied
in \cite{art2} and the expected poles for discrete momenta were
found. In this section we will
review and extend some of the results of \cite{art2} for Ward
identities. These Ward identities can be used to recursively obtain the
correlation functions.

The recent developements revealing the $W_{\infty}$ symmetries
indicate however that this is not the whole story. One should also
consider correlation functions involving powers of the conjugate
momentum. In this section we will
show the existence of this
$W_{\infty}$ symmetry which greatly will simplify the subsequent
calculations.

Ward identities for correlation functions are in general obtained by changes of
variables in the path integral. Examples of such Ward identities were obtained
in \cite{art2} from simple coordinate changes in the matrix eigenvalue.
They can be
thought of as generated by commutators, or classically i.e. on the sphere,
Poisson bracketts with $p\lambda ^{m}$.
They obviously obey a Virasoro algebra which is part of a $W_{\infty}$
algebra generated by all monomials $p^{n} \lambda ^{m}$. We may also introduce
time dependence and consider generators with certain momenta $q$, i.e.
$p^{n} \lambda ^{m} e^{iqt}$.
Let us as an example make a $p \lambda ^{k} e^{iqt}$ variation of the
one puncture function. The two puncture function is schematically given by
\begin{equation}
<PP> = {\rm Im} \int_{0}^{\infty} dT \int d\lambda  G(\lambda , \lambda ;T)
\end{equation}
where the calculation is done at the Fermi surface. We
will shift its energy to zero, hence putting the Fermi energy as a
constant term in the
potential. $G$ is the path integral given by, in Euclidean time,
\begin{equation}
G(\lambda _{1},\lambda _{2}) = \int _{\lambda _{1}}^{\lambda _{2}}
[dp d\lambda] e^{-\beta \int dt(p\dot{\lambda }-\frac{1}{2}p^{2}+U(\lambda
))} \label{ban}
\end{equation}
where $U(\lambda)=\sum _{p} t_{p} \lambda ^{p}$ is the potential.
The variation of the partition function would have involved a sum over
all states in the Fermi sea up to the Fermi level. By  inserting a
puncture, i.e. taking a derivative with respect to the Fermi energy, we
restrict ourselves to the Fermi surface. Next we perform the variation
of $<P>$. The measure, as given by (\ref{ban}), is invariant under the change
of
variables. (Clearly we are not supposed to differentiate with respect
to $t$ when changing variables in the measure.)
The change in the action give rise to the following identity
among two point functions:
\begin{equation}
< (\int dt (iq \lambda ^{k}p +k\lambda ^{k-1}p^{2}+
\frac{\partial U}{\partial \lambda} \lambda ^{k})e^{iqt}) T>=0
\end{equation}
The puncture is now a tachyon, T,  carrying away the momentum.
The piece with a single $p$ is evaluated by integrating over $p$ in the
path integral obtaining a $\dot{\lambda }$ which then is partially
integrated. We then switch to a Hamiltonian formulation, remembering
that we should use Weyl ordering.
We finally obtain:
\begin{eqnarray}
-\frac{1}{2}k(k-1)(k-2)<O_{k-3}T>_{q,g-1}+    \label{rec}
\sum _{p} (2k+p)t_{p}<O_{p+k-1}T>_{q,g} \\ \nonumber
+ \frac{q^{2}}{1+k}<O_{k+1}T>_{q,g} =0
\end{eqnarray}
where $q$ indicates momentum and $g$ genus. We have introduced the
notation $O_{k}$ for $\lambda ^{k}$.
The different genus for the
first term is due to a $p$, $\lambda$ commutator which arises when we
want to evaluate the $p^{2}$ against the wave function of the Fermi
surface. This gives an $\hbar$, which is the same as the
genus coupling constant.
Following \cite{mo} we may define the loop operator given by
\begin{equation}
w(l)=e^{l\lambda } \label{loop}
\end{equation}
It corresponds to cutting out a hole in the surface with a boundary of
length $l$. The reason is as follows. If we insert a power $n$ of the
original matrix eigenvalue $m$ on the surface this creates a little hole,
the length of the boundary being proportional to $n$ (the number of legs) and
the lattice spacing $a$. For
fixed $n$ the length clearly shrinks to zero in the double scaling
limit. To get a finite length we must also take $n$ to infinity.
Introducing $\lambda$ as $m$ expanded around the top of the potential,
we find in the double scaling limit
\begin{equation}
m^{n} \sim (1+a\lambda )^{l/a} \rightarrow e^{l\lambda}
\end{equation}
We may then Fourier transform to obtain a differential equation in the
loop length. We get
\begin{equation}
[\sum _{p} t_{p} (l^{2} \frac{\partial ^{p}}{\partial l ^{p}}
+\frac{p}{2}l\frac{\partial
^{p-1}}{\partial l^{p-1}})+t_{0}l^{2} - l^{4}
+q^{2}]<w(l)T>=0 \label{WdW}
\end{equation}
where the third term are of order one higher in the string coupling and
do not contribute on the sphere. $t_{0}$ is the Fermi energy $\mu$ with
the appropriate number of $\beta$'s absorbed.
In the case of the usual harmonic oscillator potential, where
$t_{0}=\beta \mu$, the
resulting equation is in fact the Wheeler de  Witt equation obtained in
\cite{mo}
with a more indirect matrix model method. On the sphere this is one of
the most striking verifications of the equivalence of the Liouville and
matrix models. At zero momentum the
Wheeler de Witt equation is
just the Fourier transform of the Gelfand-Diiki equation for the resolvent
for the Schr\"{o}dinger operator. This was the way in which the zero momentum
version of (\ref{rec}) was derived in \cite{art2}. If we want to be
careful, see section 5, we need to rescale $\lambda$ by $\sqrt{-t_{2}}$
to get a dimensionless $l$. This is needed for the exact correspondence
between the above result and the mini superspace canonical quantization
of Liouville theory. Recall that, \cite{new}, $t_{2}=-\frac{1}{2\alpha
'}$.
{}From (\ref{WdW}) one might try to draw some
conclusions about the Liouville theory correspondence to the more
general potentials above. Clearly the last term, which corresponds to
the matter piece, does not change while we change potential. Instead it
is the piece which would be expected to arise from a canonically
quantized kinetical term for the Liouville mode which gets modified.
Hence one is lead to the conclusion that these more general models
(however with $p$ independent potentials)
may correspond not to modifications of the matter theory but rather to
different theories for the Liouville part. This is also consistent with
the point of view for which this paper will argue, that the special
states must be represented using both $\lambda$'s and $p$'s, not just
the $\lambda$'s.

In principle all correlation functions may be calculated with the help
of Ward identities derived in this way. However, it is more convenient
to make use of the large set of symmetries in the theory.
As shown by Witten in
\cite{wit}
we may change
basis to $(p-\lambda )^{n} (p+\lambda )^{m}$ and, for certain time dependence,
obtain transformations which leave the action invariant. These
transformations are generated by, in Minkowsky time,
\begin{equation}
W^{r,s} = (p+\lambda )^{r} (p- \lambda )^{s} e^{(r-s)t} \label{w}
\end{equation}
Again we get the $W_{\infty}$ algebra
\begin{equation}
\{W^{r_{1},s_{1}},W^{r_{2},s_{2}}\}=(r_{1}s_{2}-r_{2}s_{1})W^{r_{1}+r_{2}-1,
s_{1}+s_{2}-1} \label{alg}
\end{equation}
generated, classically, by the Poisson bracketts.
For a general momentum $q$ in (\ref{w}) we find when acting on the Minkowsky
action
$S=\int (p \dot{\lambda} -\frac{1}{2} (p^{2}-\lambda ^{2}))$
\begin{equation}
\{W^{r,s},S\}=(r-s-q)W^{r,s} \label{svar}
\end{equation}
Hence a symmetry for appropriate discrete values of imaginary momentum. This is
equivalent to saying that
$W=W(p,\lambda,t)$ is
a solution of
\begin{equation}
\frac{dW}{dt} = \frac{\partial W}{\partial t} + \{H,W\} =0
\end{equation}
Expressed in terms of the initial conditions, $p_{0}$ and $\lambda
_{0}$, we have
$W= W(p_{0}, \lambda _{0})$, i.e. any time
independent function of the initial conditions. The generators (\ref{w}) are
then simply obtained through evolution in time.
Hence the transformations can be understood as time independent canonical
transformations of the initial conditions.

A more convenient way of labeling the generators is through their
$SU(2)$ quantum numbers $J=(r+s)/2$ and $m=(r-s)/2$.
With the definition
\begin{equation}
W_{J,m} = (p+\lambda )^{J+m}(p-\lambda )^{J-m} e^{2mt} \label{w2}
\end{equation}
one gets
\begin{equation}
\{W_{J_{1},m_{1}},W_{J_{2},m_{2}}\}=2(m_{1}J_{2}-m_{2}J_{1})
W_{J_{1}+J_{2}-1,m_{1}+m_{2}} \label{alg2}
\end{equation}
In Euclidean time, which is what we will be using, one should take
$p \rightarrow ip$ and $t \rightarrow it$ in (\ref{w2}). There is also
an extra $i$ in the structure constant of (\ref{alg2}) and the
eigenvalue of (\ref{svar}).

\section{Matrix Model Correlation Functions}

In this section we will calculate correlation
functions involving the operators $W$ as defined above. The $W_{\infty}$
symmetries will help us organize the Ward
identities. As an example, let us start with the two point function.
Using
\begin{equation}
<PP>=\frac{1}{\pi} {\rm Im} \sum _{n=0}^{\infty} \frac{1}{E_{n}+t_{0}}
\end{equation}
and
simple perturbation theory we get:
\begin{equation}
<W_{1}W_{2}P>=\frac{1}{\pi}{\rm Im} \sum _{timeord} \sum _{n,k}
\frac{<n \mid W_{1} \mid k>
<k \mid W_{2}
\mid n>}{E_{n}+t_{0}}\frac{1}{ip_{1}+E_{k}-E_{n}}
\end{equation}
Since the $W$'s are of the form (\ref{w}), continued to Eucledian time,
they are simply raising or lowering
operators in the inverted harmonic oscillator.
This means that only a few of the matrix elements are
nonzero. Since $W$ raises by $2m=r-s$ we get
\begin{equation}
<W_{1}W_{2}P>=\frac{1}{\pi} {\rm Im} \sum _{n} \frac{<n\mid [W_{1},W_{2}] \mid
n>}{E_{n}+t_{0}}\frac{1}{i(p_{1}-2m_{1})} \label{kom2}
\end{equation}
We have reduced the two point function to a one point function using the
commutation relations. If we restrict ourselves to the sphere, use
the algebra given by (\ref{alg}) (with an extra $i$ in the structure
constant for Eucledian time) and directly calculate the one point
function we get
\begin{equation}
<W_{1}W_{2}>=\frac{2(m_{1}J_{2}-m_{2}J_{1})}{2m_{1}-p_{1}}\frac{1}{\pi}
\frac{\mu
^{J_{1}+J_{2}}}{J_{1}+J_{2}} \mid \log \mu \mid 2^{J_{1}+J_{2}-1}\label{2p}
\end{equation}
or equivalently
\begin{equation}
<W_{1}W_{2}>=\frac{m_{1}}{2m_{1}-p_{1}}\frac{1}{\pi} \mu ^{J_{1}+J_{2}}
\mid \log \mu \mid 2^{J_{1}+J_{2}}\label{2pp}
\end{equation}

The simplest way to obtain the one point function on the sphere is to use
the classical Fermi liquid picture introduced by Polchinski \cite{pol}.
We simply need to
do the phase space integral:
\begin{equation}
<W>=\int dp d\lambda W(p,\lambda )
\end{equation}
This was also discussed in \cite{wad}.
To make everything well defined we need however to introduce an extra puncture,
i.e. take a derivative with respect to the cosmological constant. Doing that
the integral over the whole Fermi sea becomes just an integral over the Fermi
surface:
\begin{equation}
<WP>=\oint W(p, \lambda )
\end{equation}
The integral is to be performed along a hyperbola $\frac{p^{2}-\lambda
^{2}}{2}=\mu$. Although the answer is really infinite, we know that the
piece nonanalytic in $\mu$ is $\frac{1}{\pi} \mid \log \mu \mid$ for
$<PP>$, i.e. when we just integrate $1$. The other zero momentum
operators simply involve integrations of $(\frac{p^{2}-\lambda
^{2}}{2})^{n}=\mu
^{n}$, again constants along the Fermi surface hyperbola, so we get
\begin{equation}
<W^{n,n} P> = \frac{1}{\pi} \mu ^{n} \mid \log \mu \mid  2^{n}
\end{equation}
and from this (\ref{2p}) follows. Our conventions are such that $\alpha
' =1$. Another approach, which is convenient
when calculating correlation functions of nonzero momentum, is to
continue to the upside down oscillator where the Fermi surface is a
circle. In that case, however, we need to remember to put the Liouville
volume $\mid \log \mu \mid$ in by hand.
Parenthetically we may note how a general
correlation function may be obtained in this way. For instance the two point
function
is obtained by perturbing the hamiltonian and hence the Fermi surface by
one of the operators.
If we integrate the other operator against the change in Fermi surface we get
the correlation.

Let us now consider the more complicated case of a
three point function. Again perturbation theory gives us
\begin{equation}
\frac{1}{\pi} {\rm Im} \sum _{timeord} \sum _{n,m,k} \frac{<n \mid W_{1}
\mid m><m\mid W_{2}\mid k><k\mid W_{3}\mid n>}{E_{n} + t_{0}}
\frac{1}{ip_{1} + E_{m}-E_{n}}\frac{1}{ip_{3} +E_{n}-E_{k}}
\end{equation}
Let us make the sum over time orderings more explicit. We find
\begin{equation}
\frac{1}{\pi} {\rm Im} \sum _{n} \{
\frac{<n\mid (W_{1}W_{2}W_{3}+W_{3}W_{2}W_{1})\mid
n>}{E_{n}+t_{0}} \frac{1}{i(p_{1}-2m_{1})i(p_{3}-2m_{3})} + perm \}
\end{equation}
This may after some straightforward manipulations be rewritten as
\begin{equation}
<W_{1}W_{2}W_{3}>=\frac{<[W_{1},[W_{2},W_{3}]]>}{(p_{1} -2m_{1})(p_{3}+
2m_{1}+2m_{2})}
+\frac{<[W_{2},[W_{1},W_{3}]]>}{(p_{2}-2m_{2})(p_{3}+2m_{1}+2m_{2})}
\label{3kom}
\end{equation}
For the sphere we now use the algebra (\ref{alg2}) and the explicitly
calculated one point
function to get
\begin{eqnarray}
<W_{1}W_{2}W_{3}>=
\frac{(2m_{1}-p_{1})m_{2}m_{3}J_{1}+(2m_{2}-p_{2})m_{1}m_{3}J_{2} +
(2m_{3}-p_{3})m_{1}m_{2}J_{3}}{(2m_{1}-p_{1})(2m_{2}-p_{2})(2m_{3}-p_{3})}
\nonumber \\
\times \frac{1}{\pi} \mu ^{\sum _{n}^{N}
J_{n} -1 } \mid \log \mu \mid 2^{\sum_{n=1}^{N} J_{n}} \label{strul}
\end{eqnarray}

The general higher point function can be obtained recursively from the
three point by use of (\ref{alg}) and (\ref{svar}). To get the $N$
point function with an additional operator $W_{N}$ we vary the $N-1$
point function with $W_{N}$ knowing that the total variation is zero.
The variation consists of two terms. One from varying the action as
given by (\ref{svar}) and a sum of terms from varying the other
operators as given by (\ref{alg2}). Each of the terms in this sum is
obtained by shifting $p_{i} \rightarrow p_{i}+p_{N}$, $m_{i} \rightarrow
m_{i} +m_{N}$ and $J_{i} \rightarrow J_{i}+J_{N+1} -1$. We also need to
multiply with the Clebsch-Gordan coefficient $2(J_{i}m_{N}-J_{N}m_{i})$.
It is an easy exercise to check that (25) is obtained by applying
this procedure to (\ref{2p}) or (\ref{2pp}).
There is one subtlety in the variational
procedure which should be noted. The insertion of an operator $W$ in the
pathintegral does not involve only the operator itself, but also a delta
function for its position in eigenvalue space. In general the delta
function will also contribute to the variation. Luckily its contribution
will be zero by invariance properties for the $W_{\infty}$ generators.

Using this method one can write down several different recursion
relations. One simple example is:
\begin{equation}
<T_{J,J}W_{J_{1},m_{1}}\prod _{i=2}^{N} T_{J_{i},J_{i}}>=
\frac{2J(J_{1}-m_{1})}{2J-p} <W_{J+J_{1}-1,J+m_{1}}\prod_{i=2}^{N}
T_{J_{i},J_{i}}>  \label{rek}
\end{equation}
We will come back to this relation later, when we compare with the Liouville
model results.

Rather than considering these general expressions, let us look at a
couple of important examples where the form of the general $N$ point
function is particularly simple.

The first example is the $N$ point function of special tachyons. It is
given by
\begin{equation}
<\prod _{n=1}^{N} T_{n}> = \frac{2\prod _{n=1}^{N} J_{n}}{\prod
_{n=2}^{N}(2J_{n}-p_{n})}\frac{1}{\pi} \frac{d^{N-3}}{d\mu ^{N-3}} \mu ^{\sum
_{n=1}^{N} J_{n} -1} \mid \log \mu \mid 2^{\sum_{n=1}^{N} J_{n}} \label{tach}
\end{equation}
The quantum numbers have been chosen as
$m_{n}=J_{n}$ for $n>1$ and
$m_{1}=-J_{1}$.
This is just the pole part of the general tachyon correlation function as
computed both in the matrix model \cite{grkl} and in the Liouville theory
\cite{grkl,frku}, up to
a factorized normalization factor. The proof is by varying the three
point. We can not just vary the three point tachyon correlation
function, since some of the $J$'s are really $m$'s in disguise and $J$
and $m$ vary differently. Instead we start with the general three point
and make an arbitrary number of tachyon variations. A simplification is
that we at each step only have to vary the single negative chirality
tachyon. It is only from there were we will get a nonzero
Clebsch-Gordan
coefficient. Following the
prescription above, performing $N-3$ variations we get

\newpage

\begin{eqnarray}
<\prod _{n=1}^{N} T_{n} > =[(2m_{1}-p_{1} +\sum _{i=4}^{N}
(2m_{i}-p_{i}))m_{2}m_{3}(J_{1}+\sum _{i=4}^{N}J_{i} -N+3) \nonumber \\
+(2m_{2}-p_{2})(m_{1}+\sum _{i=4}^{N} m_{i})m_{3}J_{2}
+(2m_{3}-p_{3})(m_{1}+\sum _{i=4}^{N} m_{i})m_{2}J_{3}] \nonumber \\
\times \frac{\prod _{i=4}^{N} (J_{i}(m_{1}+\sum
_{j=i+1}^{N}m_{j})-(J_{1}+\sum_{j=i+1}^{N}J_{j}-N+i)m_{i})}
{(2m_{1}-p_{1}+\sum _{i=4}^{N} (2m_{i}-p_{i}))\prod_{n=2}^{N}(2m_{n}-p_{n})}
\frac{1}{\pi} \mu ^{\sum _{n=1}^{N} J_{n} -N+2} \mid \log \mu \mid
2^{\sum_{n=1}^{N} J_{n}}
\label{tok}
\end{eqnarray}
The product in the denominator is the product of all the Clebsch-Gordan
coefficients of the variations. Note that each get shifted by the
successive variations. By the use of momentum conservation and evaluating
the $m$'s as $J$'s, the formula (\ref{tach}) is proved. This derivation
shows how the combinatorical factor from the $\mu$ derivatives is a
consequence of the $W_{\infty}$ symmetry.

If we want to consider the zero momentum operators,
we have to be careful. The Clebsch-Gordan coefficients are zero in this case
but these zeroes cancel precisely the momentum poles and leave a finite result.
Also, we need to consider both signs of the $m$ quantum number when we take the
$m \rightarrow 0$ limit. This gives a necessary extra factor of two.
We get
\begin{equation}
<\prod _{n=1}^{N} W_{n}>=\sum _{n=1}^{N} J_{n} \frac{1}{\pi}
\frac{d^{N-3}}{d \mu
^{N-3}} \mu ^{\sum _{n=1}^{N} J_{n} -1} \mid \log \mu \mid
2^{\sum_{n=1}^{N} J_{n}}
\end{equation}
We will use induction for the proof. We find
\begin{eqnarray}
<\prod _{n=1}^{N+1}W_{N}> = \sum_{k}2(J_{N+1}m_{k}-J_{k}m_{N+1}) \frac{\sum
_{n=1}^{N+1}J_{n}-1}{2m_{N+1}-p_{N+1}} \nonumber \\ \times \frac{1}{\pi}
\frac{d^{N-3}}{d \mu
^{N-3}} \mu ^{\sum _{n=1}^{N+1} J_{n} -2} \mid \log \mu \mid
2^{\sum_{n=1}^{N+1} J_{n} -1}
+(m_{N+1} \rightarrow -m_{N+1})
\end{eqnarray}
If we then put $p_{N+1}=0$ and use that the sum of all $m$'s must be
zero the result follows.
This can also be checked by an explicit phase space calculation.

Given these expressions we may check the correlation functions calculated in
\cite{art2}
and independently in \cite{mo}. These were correlation functions of pure powers
of the
matrix eigenvalue without any momentum powers. In terms of the W's they are
given by
\begin{equation}
O_{n}=\sum _{k=0}^{n} \left(\begin{array}{c}
                               n \\ k
                            \end{array} \right)
(-1)^{k} W^{n,k-n} \frac{1}{2^{n}}
\end{equation}
{}From this it follows that the two point function is given by
\begin{equation}
<O_{n}O_{m}>_{q} = \frac{1}{2^{n+m}}\sum _{k=0}^{n} \sum _{l=0}^{m}
\left(\begin{array}{c}
n \\ k
\end{array} \right)
\left(\begin{array}{c}
m \\ l
\end{array}\right)
(-1)^{k+l} <W^{n-k,k}_{q}W^{m-l,l}_{-q}>
\end{equation}
Using (\ref{2pp}) and some simple algebra we find
\begin{equation}
\frac{1}{2^{\frac{1}{2}(n+m)}} \frac{1}{\pi}
\mu ^{\frac{1}{2}(n+m)}\mid \log \mu \mid \sum _{k=0}^{n}
\left(\begin{array}{c}
n \\ k
\end{array}\right)
\left(\begin{array}{c}
m \\ \frac{n+m}{2} -k
\end{array}\right)
\frac{4(\frac{n}{2} -k)^{2}}{4(\frac{n}{2}-k)^{2}-q^{2}}
\end{equation}
In precise agreement with \cite{art2} recalling our convention $\alpha
'=1$.
We can now understand why the $O$ operators gave correlation functions with
sets
of poles and were, depending on momentum, capable of exciting several special
states \cite{art2}. They were, in fact, linear combinations of all
special operators of
a given gravitational dimension i.e. spin $J$. The above construction with the
generators (\ref{w}) of the $W_{\infty}$ disentangles the correlation
functions.
This means that the matrix model operators to be identified with the
Liouville model special states are those defined in (\ref{w2}).

Finally let us consider the meaning of the momentum poles.
As emphasized in \cite{klre} we should not treat the poles
in (25) and the $\mid \log \mu  \mid$ asymmetrically since the
source of the $\mid \log \mu \mid$ is also a momentum pole.
In fact, all the poles should be thought of as cut of by $\mid \log
\mu \mid$.
A general $N$ point function (without zero momentum operators) would
then have $\mid \log \mu \mid ^{N}$.
This proliferation of logaritms was also noted in \cite{frku}.

\section{Higher Genus}

We have so far basically just treated the sphere, which means, in the
matrix model, that we have been working at the classical level. The
$W_{\infty}$ has been generated by Poisson bracketts. Nothing can
however stop us from considering the full quantum theory, i.e. all
genus. It is just a matter of algebra to compute for instance the two
or three
point functions using (\ref{kom2}) and (\ref{3kom}) respectively.

More interestingly, the algebra changes at the quantum level. There is a
deformation with $\hbar$, the genus coupling, as parameter when we use
commutators instead of Poisson bracketts. If we define our $W$'s using
Weyl ordering, which is natural from the pathintegral point of view, the
algebra may be conveniently represented using the Moyal bracketts \cite{moy}
\begin{equation}
\{ W_{1},W_{2} \} _{M} = \frac{2}{\hbar} \sin \frac{\hbar}{2}
(\frac{\partial}{\partial
p_{1}} \frac{\partial}{\partial \lambda _{2}} -
\frac{\partial}{\partial
p_{2}} \frac{\partial}{\partial \lambda _{1}}) W_{1}W_{2}
\end{equation}
which is a deformation of the usual Poisson brackett. From this we might
conclude that also the Liouville theory operator product expansions
should receive higher genus corrections. Presumably from handles getting
caught inside the contour integrals defining the operator product
expansions.

Another way to exhibit the quantum deformation is through a generalized
loop operator \cite{wad}, instead of (\ref{loop}) we introduce
\begin{equation}
w(k,l)=e^{kp+l\lambda}
\end{equation}
where $p$ and $\lambda$ are the conjugate variables. It is a very old
result, \cite{neu}, that these operators obey the algebra
\begin{equation}
[w(k_{1},l_{1}),w(k_{2},l_{2})]=\frac{1}{\hbar} \sin
\hbar (k_{1}l_{2}-l_{1}k_{1}) w(k_{1}+k_{2},l_{1}+l_{2}) \label{sun}
\end{equation}
with $\hbar \rightarrow 0$ giving back a $W_{\infty}$. Interestingly it can
be shown \cite{bars} that (\ref{sun}) is a representation of $SU(N)$ with
$\hbar
=1/N$.
This is
reminiscent of the original unitary symmetry of the matrix model.

Let us give an explicit example of a two point function to all
genus. We choose the correlator between spin $J=3/2$, $m=1/2$ and
$J=3/2$, $m=-1/2$. To do that we need to calculate
$<W_{2,0}P>$. This is easy. We have
\begin{equation}
<W_{2,0}P>=\frac{1}{\pi} {\rm Im} \sum _{n=0}^{\infty}
\frac{(2E_{n})^{2}+1}{E_{n}+t_{0}}
\end{equation}
The extra term $+1$ comes from Weyl ordering.
If we keep only terms nonanalytic in $t_{0}$ this reduces to
\begin{equation}
<W_{2,0}P>=(4t_{0}^{2} +1) <PP> \label{1p}
\end{equation}
To evaluate our two point we use the Moyal brackett to calculate
\begin{equation}
[W_{3/2,1/2},W_{3/2,-1/2}]=6W_{2,0} -4\hbar ^{2}W_{0,0}
\end{equation}
(\ref{kom2}) and (\ref{1p}) then finally give
\begin{equation}
<W_{3/2,1/2}W_{3/2,-1/2}P>=(48t_{0}^{2}+4)<PP>\frac{1/2}{1-p}
\end{equation}
The same procedure may be used to calculate arbitrary correlation
functions.

\section{Comparison with Liouville Theory}

We would like to understand the $W_{\infty}$ structure from the
Liouville theory point of view. As shown in \cite{wit} and by more direct
methods in \cite{klpol}, we indeed have the same algebraic structure present.
Therefore one would expect the comparison of the Liouville and the
matrix model to be straightforward. As we will see, the
situation is more subtle.
Let us first consider a very special case which
also give us the opportunity to clarify some important points.

There are some very simple examples of correlation functions easily
computable just using Liouville notions and no matrix model techniques.
These are correlation functions involving the dilaton. We will in fact
be able to obtain some results to all genus simply from dimensional
arguments. Consider the Liouville partition function (or space time free
energy)
\begin{equation}
E(\Delta ) = \lim _{R \rightarrow \infty} \frac{1}{R} \int DX D\phi
e^{-\int  (-t_{2} \partial X \bar{\partial} X +\partial \phi
\bar{\partial} \phi +QR\phi +\Delta e^{\alpha \phi})} \label{path}
\end{equation}
where $t_{2}=-\frac{1}{2\alpha '}$, $Q=2\sqrt{2}$, $\alpha =-\sqrt{2}$ and $R$
is
the radius of the target space for the matter field $X$.
$\Delta$ is the world sheet cosmological constant, dimensionless from
the point of view of space time. The only dimensionful quantities are
$R$ and $\alpha '$. In the noncompact case we have infact only $\alpha '$
at our disposal. From dimensional grounds and KPZ scaling we must have
\begin{equation}
E(\Delta )_{g} \sim (-t_{2})^{1/2} \Delta ^{2(1-g)} \label{E}
\end{equation}
at genus g. $E(\Delta )$ is the generator of connected amplitudes (in
space time). Let us do a Legendre transform to obtain a generating
functional for 1PI amplitudes with respect to the puncture, i.e. the
zero momentum tachyon. This means taking away any pinches. We have
\begin{equation}
E(\Delta ) = \Delta \mu - \Gamma (\mu )
\end{equation}
with
\begin{equation}
\mu = \frac{\partial E}{\partial \Delta} = (-t_{2})^{1/2} (\Delta +...)
\end{equation}
$\mu$ has the dimensions of energy. Hence
\begin{equation}
\Gamma (\mu ) \sim (-t_{2})^{\frac{1}{2}(2g-1)} \mu ^{2-2g} \label{leg}
\end{equation}
In the 1PI generator $\Gamma$ we should of course regard $\mu$ as independent
of $t_{2}$.
Since $t_{2}$ derivatives should generate dilaton insertions we find the
following 1PI amplitude relation
\begin{equation}
<O_{2}...O_{2}>_{g} = \frac{1}{(2t_{2})^{n}} \prod _{p=1}^{n} (2g+1
-2p)<>_{g}
\end{equation}
which is identical to what was obtained in \cite{art2} using the matrix model
recursion relations (generalizations of the zero momentum Wheeler de
Witt equation). As noted in \cite{art2} the dilaton one point function
involves a factor $2g-1$ rather than the expected $2g-2$. From above it
is clear that this discrepancy is simply due to including the overall
$(-t_{2})^{1/2}$ in (\ref{E}). There is a further subtlety in how the
dilaton is defined. As we have seen a pure $\lambda
^{2}$ is not what we would expect to identify with the dilaton. Instead
we should have $p^{2}-\lambda^{2}$ if we keep the algebraic structure in
mind. These are in general different in
correlation functions. We we will return to this shortly.

It is important to realize the difference between connected and 1PI
amplitudes. In the matrix model 1PI amplitudes are the natural objects,
in Liouville theory it is more common to treat the connected ones. Often
the distinction is not very clearly made. Indeed if we consider generic
nonzero momentum the difference is very easy to deal with. It
amounts to a renormalization of the cosmological constant. We can simply
replace $\Delta$ by $\mu$, \cite{klre}.
At zero momentum we must be much more careful. In this case we may have
internal puncture propagators, i.e. pinches. Some examples of this were
obtained in \cite{art2}. This will turn out to be important later on.

Already at this point we may find traces of the $W_{\infty}$ structure.
In fact, the seemingly innocent representation of the puncture and the
dilaton as $\mu$ and $t_{2}$ derivatives respectively is a reflection
of the $W_{\infty}$. Let us give a formal argument for this. First the
puncture. Write the
$SU(2)$ quantum numbers of the puncture as $J$ and $m$ which both will
be taken to zero. Choose one of the operators in (25) to be a
puncture. We get
\begin{eqnarray}
<W_{1}W_{2}P>=(J_{1}+J_{2})\frac{m_{1}}{2m_{1}-p_{1}} \frac{1}{\pi} \mu
^{J_{1}+J_{2}-1} \mid \log \mu \mid
2^{J_{1}+J_{2}}
\label{puder}
\end{eqnarray}
which by comparing with the two point function (\ref{2p}) shows how the
puncture is represented as a $\mu$ derivative.
The case of the dilaton is equally simple.
Proceeding as
above we find
\begin{equation}
<W_{1}W_{2}D>=[(J_{1}+J_{2})\frac{m_{1}}{2m_{1}-p_{1}} +
\frac{2m_{1}^{2}}{(2m_{1}-p_{1})^{2}}] \frac{1}{\pi}
\mu ^{J_{1}+J_{2}} \mid \log \mu
\mid
2^{J_{1}+J_{2}+1} \label{3pD}
\end{equation}
If we introduce explicit $t_{2}$'s in the two point function we can
write it as
\begin{equation}
<W_{1}W_{2}>=\frac{m_{1}}{2m_{1}-p_{1}/(-2t_{2})^{1/2}} \frac{1}{\pi}
(\frac{\mu}{(-2t_{2})^{1/2}})^{J_{1}+J_{2}} \mid \log \mu \mid
2^{J_{1}+J_{2}} \label{2t2}
\end{equation}
Taking a $t_{2}^{1/2}$ derivative we indeed reproduce (\ref{3pD}). We
must now return to the issue of how
precisely the dilaton is defined. Recall
the original matrix model action:
\begin{equation}
\beta \int dt [p\dot{\lambda} -\frac{1}{2}p^{2}-t_{2} \lambda ^{2}]
\end{equation}
with $\beta$ dimensionless, $t_{2}$ having the dimension of energy squared
and $p^{2}$ and $\lambda^{2}$ the dimensions of energy and one over energy
respectively. To obtain (\ref{2t2}) as a
generating functional for dilaton insertions with the above definition
of the dilaton we should rescale $\lambda$ and $p$ to make them
dimensionless. We find
\begin{equation}
\beta \int dt [p\dot{\lambda}-\frac{1}{\sqrt{2}}(-t_{2})^{1/2} (p^{2}-\lambda
^{2})]
\end{equation}
Hence the matrix model dilaton should be represented by $(-t_{2})^{1/2}$
derivatives. This is the rescaling eluded to in section 2 in the context
of the Wheeler de Witt equation.

To obtain the general special operator correlation function in the
Liouville theory one would like to use the group theoretic
information provided by
the $W_{\infty}$ or, for given spin $J$, the $SU(2)$ symmetry.
The states in the Liouville theory are given by combinations
$W(z)\bar{W}(\bar{z})$ of the Liouville theory version of the special
states $W(z)$. Given this it is tempting to
believe that we have a representation of a
$W_{\infty} \times W_{\infty}$ symmetry
(left times right). This is however in general not correct.
In the uncompactified
case the left and the right moving states must be the same.
The symmetry group is
broken down to just the diagonal subgroup.
This is achieved in two steps. First the gravitational dressing must be
the same for left and right, otherwise we would be unable to screen
using the cosmological constant which treats left and right in the same
way. This means that we always must have the same spin $J$ for left and
right. We get a reduction to the diagonal of the piece transverse to
$SU(2) \times SU(2)$. This is true even for the compactified case.
If we in addition are considering the
uncompactified case, the left and right moving momenta must be the same
and hence the $m$ quantum numbers. Consequently we just have a
representation of the diagonal $W_{\infty}$.
This is in precise agreement with the
matrix model, where we indeed only see one $W_{\infty}$. There is
however an apparent paradox here. If we would use the free field
contractions in computing the correlation functions the results would
seem to disagree since from this point of view left and right are still
independent. We will return to this important point further on, and
discover that there in fact seems to be no contradiction.

The symmetry may then be used to determine all correlation functions
given the special tachyon correlation functions which may be computed
using other means. The reason is that all $J$ and $m$ dependence of any
correlation function is given by some combination of Clebsch-Gordan
coefficients. For given $J$'s we need the Clebsch-Gordan coefficients of
$SU(2)$, the $3j$ symbols, to get the $m$ dependence.
In fact we have
already seen the agreement for the tachyon correlation functions and
if we believe that the group theoretic structure is the same in
the matrix model and
in the Liouville theory, we know that the expressions obtained in the
matrix model must agree with Liouville theory.
To be more explicit let us however look
at an example, the three point function, to see how the invariance
properties determine the correlation functions. The three point function
is
obtained by considering
coupling $(J_{1},m_{1}), (J_{2},m_{2})$ and $(J_{3},m_{3})$ (with
$m_{1}+m_{2}+m_{3}=0$) to $(J_{1}+J_{2}+J_{3}-2,0)$. A complication is
that there are in general several different channels to sum over. This
is true already for the three point function. The reason is that we
really should think of the three point function as a four point
function. The fourth leg carries the excess Liouville momentum, i.e. $J$
quantum number, into the vacuum. This is a consequence of the non
conservation of Liouville momentum. Let us use the tachyon three
point function for normalization. It is given by
\begin{equation}
<T_{1}T_{2}T_{3}>=\frac{J_{1}J_{2}J_{3}}{(2m_{2}-p_{2})(2m_{3}-p_{3})}
\frac{1}{\pi}
\mu
^{J_{1}+J_{2}+J_{3}-1}\mid \log \mu \mid 2^{J_{1}+J_{2}+J_{3}+1} \label{t3}
\end{equation}
where we have kept the normalization choice of (\ref{tach}).
Tachyons 2 and 3 are of positive chirality while tachyon 1 has negative
chirality.
There are two possible channels corresponding to either $p_{2}=m_{2}$ or
$p_{3}=m_{3}$, i.e. 1 and 2 coming together or 1 and 3 coming together.
The group theoretic factor in each case is simply proportional to a product of
$3j$ symbols. One for each vertex. For the 1-2 channel:
\begin{equation}
\left(\begin{array}{c}
         J_{1} J_{2} J \\
         m_{1} m_{2} m_{3}
      \end{array} \right)
\left(\begin{array}{c}
         J J_{3} J^{\prime} \\
         -m_{3} m_{3} 0
      \end{array} \right)
\end{equation}
where $J = J_{1}+J_{2}-1$ and $J^{\prime}=J_{1}+J_{2}+J_{3}-2$.
Just retaining the $m$ dependence and adding the two channels we find
\begin{eqnarray}
(2m_{2}-p_{2})(J_{1}m_{2}-J_{2}m_{1})m_{3}+(2m_{3}-p_{3})(J_{1}m_{3}-J_{3}m_{1})
m_{2}= \nonumber \\
-(2m_{1}-p_{1})J_{1}m_{2}m_{3}
-(2m_{2}-p_{2})J_{2}m_{1}m_{3}
-(2m_{3}-p_{3})J_{3}m_{1}m_{2}
\end{eqnarray}
which agrees with (25) after using (\ref{t3}) to fix the normalization and
$J$ dependence. This should come as no surprise since the calculations are
almost identical.

Another convenient way to obtain more general correlation functions is
through factorization.
This is really already implicit in
our previous calculations. In fact, if we look at (\ref{tok}) we see the
complete factorization of the tachyon correlation function into a
product of three point functions, each given by a $3j$ symbol, times a
single zero momentum one point function. This last piece represents the
extra leg in any correlation function which absorbs excess Liouville
momenta. One may note that these three point functions in fact involve
states of the wrong dressing. This was also pointed out in \cite{tan}.
Strictly speaking the expression in (\ref{tok}) is just for one
channel, the one where 1 fuses with 2 then with 3 etc. All channels give
however identical contributions and can not be distinguished. Clearly
the tachyon correlation function is consistent with the single
$W_{\infty}$ factorization result.

As has been remarked, this seems to be in contradiction with what to expect
from the naiive free field calculations in Liouville theory. From such a
calculation you would expect to get a different result, all
Clebsch-Gordan coefficients squared, one from the left and one from the
right. We will however show that the results in the end turn out to be
consistent. Let us begin by considering the
tachyon correlation function as computed in \cite{frku}.
As we have seen the
result is in complete agreement with the matrix model results. On the
other hand we have seen how the matrix model organizes its correlation
functions using a {\em single} $W_{\infty}$. Let us consider the
Liouville calculation more carefully.
The result of \cite{frku}
is obtained
through arguments of
analyticity and symmetry. In particular the by now well known factorized
product of gamma functions is found \cite{grkl}
with a certain unknown coefficient
independent of the particular momenta. This coefficient is then
determined by sending all the momenta, except three, to zero. This
reduces the expression to a three point function with $N-3$ extra
punctures. Since the three point function is possible to evaluate
directly, the
general result follows. The extra $N-3$ punctures is simply represented
as $\mu$ derivatives. This is the Liouville derivation of the expression
(\ref{tach}).
The important point is that the use of a $\mu$ derivative for inserting
a puncture is a consequence of having just one $W_{\infty}$! This means
that the calculation in \cite{frku} automatically incorporates this
feature.

For the more general case with nontachyonic special states, we return to
the recursion relation (\ref{rek}). Let us redefine the fields according
to
\begin{equation}
W_{J,m}=\frac{2J}{2m-p} \tilde{W}_{J,m}
\end{equation}
Then the recursion relation takes the form
\begin{equation}
<\tilde{T}_{J,J}\tilde{W}_{J_{1},m_{1}}\prod _{i=2}^{N}
\tilde{T}_{J_{i},J_{i}}>=
\frac{(J_{1}-m_{1})(J+J_{1}-1)}{J_{1}}
<\tilde{W}_{J+J_{1}-1,J+m_{1}}\prod_{i=2}^{N}
\tilde{T}_{J_{i},J_{i}}>  \label{tjohej}
\end{equation}
In \cite{kit} these very same recursion relations were obtained in the case
$J=m=1/2$ using Liouville methods. The coefficient in front of the right
hand side  were shown to be of the form $(2J_{1}-1)C^{2}$, where $C$
stands for the appropriate Clebsch-Gordan coefficient. The first factor
comes from comparing with the purely tachyonic case where the answer is
obtained from a simple Veneziano like integral.
To see the agreement one uses the
Clebsch-Gordan coefficients of the special operator algebra as
obtained in \cite{klpol}.
\begin{equation} \label{cg}
C_{J_{1},m_{1},J_{2},m_{2}}^{J_{3},m_{3}} = \frac{A(J_{3},m_{3})}
{A(J_{1},m_{1})A(J_{2},m_{2})}(J_{1}m_{2}-J_{2}m_{1})
\end{equation}
where
\begin{equation}
A(J,m)=-\frac{1}{2}[(2J)!(J+m)!(J-m)!]^{1/2}
\end{equation}
At $J=m=1/2$ one finds $C^{2} = \frac{J_{1}-m_{1}}{2J_{1}}$ which then
leads to (\ref{tjohej}).
This is an important check on the equivalence between the Liouville and
matrix model approaches. An everywhere present difficulty in these
comparisons is, however, the fact that we are really sitting right on
the momentum
poles. Clearly one needs to carefully regularize all expressions.

Let us give some further illustrations in the case
of puncture
and dilaton
insertions.
We begin with the puncture. Starting with a general correlation
function and inserting a puncture does not change the Veneziano like
integral which has to be calculated. When we insert a puncture we also
must remove one of the screening insertions. The only thing which
changes is the zero mode part of the calculation.
We recall the result
\begin{equation}
\int d\phi e^{m\phi -\Delta e^{-\phi}} \sim \Gamma (-m) \Delta ^{m}
\end{equation}
If we start with
$\Gamma (-m) \mu ^{m}$ we end up with $\Gamma (-m+1) \mu
^{m-1}=-m\Gamma(-m) \mu ^{m-1}$ when we remove a screening insertion.
Comparing with (\ref{puder}) this shows the origin of the $W_{\infty}$
related factor in the tachyon correlation function. It is a consequence
of the changing number of puncture screening operators needed.
For the tachyons the issue of connected or 1PI amplitudes
is trivial for our case with just one tachyon of differing chirality.
There can't be any internal punctures just from kinematics. This is no
longer the case when we turn to the dilaton.
The crucial point is that the dilaton can be represented
as a $t_{2}$ derivative. Usually this is precisely equivalent to using
the ordinary free field contractions giving Veneziano like correlation
functions. Inserting a dilaton in some tachyon correlation function
means taking derivatives with respect to $t_{2}$ (i.e. $1/\alpha '$).
For dimensional reasons all tachyon momenta are accompanied by an
$t_{2}$. Without explicit $t_{2}$'s one could write
$k\frac{\partial}{\partial k}$ for the dilaton. For a dilaton insertion in a
nonzero momentum correlation function the dominating pole contribution
comes from letting the $t_{2}$ derivative act directly on the poles. All
other terms are clearly less singular. This gives the second term in
(\ref{3pD}). At zero momentum we must also consider the
dependence from the $t_{2}$'s which go together with
the $\mu$'s. The latter is a consequence of dealing with 1PI rather than
connected correlation functions. One could in fact obtain the result by
considering the explicit combinations of 1PI amplitudes into connected
ones. To be more precise, if we want to obtain the 1PI amplitude from
the connected amplitude, we must amputate external puncture legs but
also subtract of the diagrams with internal puncture propagators. In
particular we need to subtract a diagram where a puncture goes off and
converts into a dilaton. This diagram therefore involves a puncture
insertion and gives a contribution corresponding to a $\mu$ derivative.
This is then simulated by an explicit $t_{2}$ accompaning the $\mu$'s to
assure the proper subtraction. This corresponds to the first term in
(\ref{3pD}), which only becomes relevant compared to the second term
at zero momentum. Otherwise we will get
one
power less of $\mid \log \mu \mid$'s. We need not restrict ourselves to a
dilaton among special tachyons, the same reasoning works for a dilaton
inserted in a general correlation function. From these two examples we
can conclude that the 1PI nature of the correlation functions is very
important in the case of zero momentum.

\section{Conclusions}

We have investigated the structure of special operator correlation
functions in $c=1$ quantum gravity. Due to the presence of a
$W_{\infty}$ symmetry the calculations become very simple. We have
also investigated the connection between the Liouville and the matrix
model, indicating the agreement for the correlation functions.

An important point is the existence in the uncompactified Liouville as
well as matrix model formulation of c=1 of just {\em one} $W_{\infty}$.
{}From the Liouville point of view this is somewhat obscure since the
operator product expansion and its Clebsch-Gordan coefficients seem to
give a structure corresponding to {\em two} $W_{\infty}$'s. Fortunately
the final outcome of the explicit calculations are identical. Further
work is however needed to establish the full equivalence.

A part of the problem is that many of the calculations are so ill
defined.  The reason is that we are sitting right on the discrete
momentum poles. Especially in the Liouville theory this is a big
technical problem.
Often we must rely on guesswork concerning ill defined
analytical continuations. It is very doubtful if many of the results
would have been obtained correctly without knowing the answers in
advance, given by the much more powerful matrix model.

Important issues for future research are to investigate multicritical
points of matrix models with generalized potentials. It is natural to
consider even non quadratic dependence on the momentum. Such
perturbations could arise from adding the special states as we have
seen. Another important point is to identify the 'wrongly' dressed
special states in the matrix model.
Such states have negative gravitational dimensions and
are important in the context of the two dimensional black hole.

\section*{Acknowledgements}

I would like to thank David Gross and Igor Klebanov for numerous discussions.

\end{document}